\documentclass[aps,twocolumn,superscriptaddress]{revtex4-1}
\usepackage{amsmath,amssymb}
\usepackage{graphics,graphicx}
\usepackage{dcolumn,bm}
\usepackage{psfrag}
\usepackage{color}
\usepackage{multirow}
\newcommand{\cu}
{\affiliation{Department of Physics, University of Calcutta,
92 Acharya Prafulla Chandra Road, Kolkata 700009, India.}}

\newcommand{\vic}
{\affiliation{Physics department, Victoria Institution College, 78B A.P. C. Road,  Kolkata 70009, India.}}

\begin{document}

\title{
Space-time  dependence of corona virus (COVID-19) outbreak
}

\author{Kathakali Biswas}
\vic
\cu
\author{Parongama Sen}%
\cu
%\author{Purusattam Ray}%
%\imsc
%\hbni

\begin{abstract}
We analyse the data for the global corona virus (COVID-19) outbreak  using the results of a previously studied  
Susceptible-Infected-Removed (SIR)  model of epidemic spreading on Euclidean networks.  We also directly study the 
correlation of the distance from the epicenter and the number of cases. An inverse square law is seen to exist approximately.
The studies are made for China and the rest of the world separately.

\end{abstract}

\maketitle

The novel 
corona virus (COVID-19), which causes an acute respiratory disease in humans, has emerged as the latest    worldwide epidemic, having  already claimed a considerable number of lives, especially in China, from where it started in late 2019
\cite{nature,abnormal}.  
The epicenter has been  
 identified as the city of Wuhan in mainland China. 
Due to drastic  precautionary  steps  taken in China,  the disease 
has been contained  to some extent   of late, however, it has rapidly spread over to the rest of the world (ROW henceforth),
causing serious concern.
It is expected that because of the current awareness,  efficient treatment  and preventive measures in operation, the number of deaths may  be controlled soon, but long term effects on academic activities, commerce, social life, sports, tourism etc. are anticipated.

\begin{figure}
\includegraphics[width = 7cm]{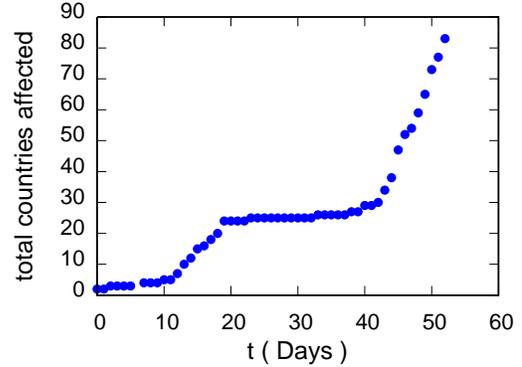}
\caption{The number of countries with corona virus patients as a function of time ($t=0$ corresponds to January 13, 2020).}
\label{fig0:countries}
\end{figure}

A considerable number of analysis of the available data of the number of cases and deaths have already been made, 
and a  few data driven models have also  been
proposed \cite{italy,chinese,newsstand,china2,crowd,delay,china3,deep,scaling,effect,artificial,visual,dynamic,trend}.  However, these works are mostly based on the data from a single country/region. In this work, the ongoing outbreak  is studied using the data in \cite{data}
where the number of total cases, number of new cases and some other relevant information  
are available for  China and other parts of the world.  The data are available in the form of daily reports (starting from January 21, 2020; the onset of the disease happened  much earlier in China) 
of new cases and total cases 
for different countries. Fig. \ref{fig0:countries} shows the number  of countries affected as a function of time (starting date January 13, 2020 for this data), the total number at present is close to 90. 
We analyse some of the data using a previously studied model for epidemic spreading on complex networks \cite{khaleq1}. 

\begin{figure}
\includegraphics[width = 7.5cm]{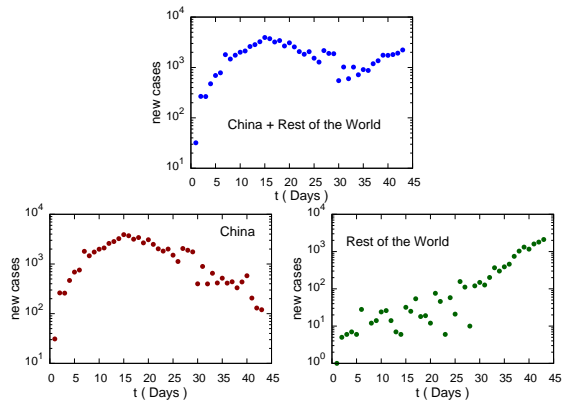}
\caption{The number of new cases versus time: upper panel shows the total number,
lower panel (left) for China only and right for rest of the world (ROW)
($t=0$ corresponds to January 21, 2020).}
\label{fig1:newcases}
\end{figure}

The affliction caused by  corona virus  belongs to the class of Susceptible-Infected-Removed (SIR) disease, where  an infected 
person either dies   or recovers, and the disease can be contracted only once. 
For the newly detected cases, SIR typically shows a peak. However, real data 
may show  multiple peaks due to delayed spreads and other, possibly demographic and geographic  reasons. 
Here also, the total number of new cases (daily)  shows a secondary growth beyond an initial  peak value (see Fig. \ref{fig1:newcases} upper panel).
That this  is  due to the later global spread beyond China is  clearly indicated  when the data for China and ROW  are plotted  separately
in the lower panel of  Fig. \ref{fig1:newcases}. 
%We first discuss the data for China where the main bulk of data are available.
The number of  newly infected cases for China shows the  feature  of a SIR type of 
disease  as it shows a   peak value followed by a decay, although not very smooth. 

The above observation shows that it is better to study the data for China and the rest of the world separately, if one uses a SIR type of model.
Since the newly infected number shows lots of fluctuations usually,
it is more convenient to  consider the  cumulative data. 
%The cumulative data  shows a saturation value  
%as expected in SIR model for China.  

In the SIR model studied on a Euclidean network \cite{khaleq1}, it was assumed that the disease can be transmitted to a nearest neighbour 
and to some random other agent who is connected with a probability decaying algebraically with the Euclidean distance 
separating them.
We attempt to fit the cumulative data using the results obtained in  \cite{khaleq1} which gave very nice agreement with the Ebola outbreak in West Africa \cite{khaleq2}. 
The transmission of corona virus  however, is much more probable than in  Ebola 
in the latter one  requires more intimate body contact to get infected. 

The cumulative data $R$ (total number of cases) for China indeed show a saturation and  are fitted to the form
obtained in \cite{khaleq1}:
\begin{equation}
R_{Ch} = A_{Ch} \exp(t/T_{Ch})/[1+ B_{Ch} \exp(t/T_{Ch})] 
\label{eq1}
\end{equation}
such that at $t  \to \infty$, a saturation value is obtained.
The value of the fitting parameters are
$ A_{Ch} \approx 2000; T_{Ch} = 5.3 \pm 0.27$ and $B_{Ch}=0.02$.
It can be seen that on day 26, there is a jump in the number (probably due to non-availability of 
quality data) and therefore the errors involved in fitting are not very small for $A_{Ch}$ and $B_{Ch}$, but definitely  less than $\sim 20\%$. 
%It may be noted here that day 0 here is  January 21, 2020, 
%from which day the data are available in \cite{data}, the actual
%onset of the disease is much earlier. 

\begin{figure}
\includegraphics[width = 7.5cm]{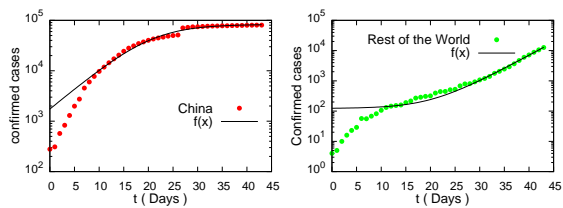}
\caption{The cumulative number of new cases versus time: 
 left panel  for China only and right for rest of the world (R0W)
($t=0$ corresponds to January 21, 2020). The best fitted functions according to the forms given in equations (\ref{eq1}) and (\ref{eq2}) are shown alongwith.}
\label{fig2:cumucase}
\end{figure}

The data for the rest of the world (ROW) 
%(order of magnitude smaller than that of China at least 
%for most part of the data)  
do not show a peak value yet for the 
newly infected cases indicating that it is still in the growing stage
 (see Fig. \ref{fig1:newcases} lower panel). 
It is not unusual to find a exponential growth initially in epidemic spreading
which would also give a exponential rise for the cumulative data.  An  
initial  exponential variation is  expected from 
 eq \ref{eq1} provided the parameter $B \ll A$ which was noted for Ebola \cite{khaleq2}.
However, we find that the present data are  better fitted with an exponential fit beyond $t=25$, in case 
one attempts to fit the entire data, the following form gives a fairly good fit: 
\begin{equation}
R_{ROW} = A_{ROW}\exp(t/T_{ROW}) + R_0
\label{eq2}
\end{equation}
with $A_{ROW} \approx 2; T_{ROW} = 4.93 \pm 0.06 $ and $R_0 \approx 122$. 
The data and the fittings for both China and ROW are shown in Fig. \ref{fig2:cumucase}.

%In case one plots the total data for the entire world, one gets a behaviour like China as the number of cases etc are order of magnitude less
%from the rst of the world. 

\begin{figure}
\includegraphics[width = 7cm]{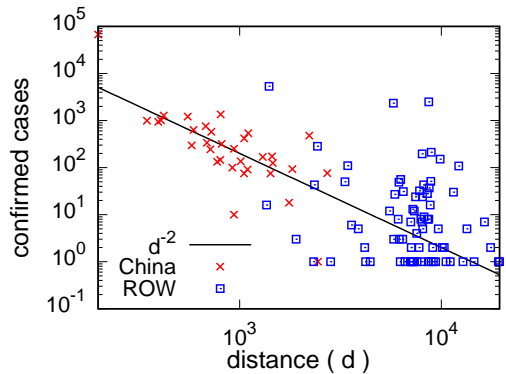}
\caption{The cumulative number of new cases  recorded in a place at a distance $d$ Km  from the epicenter Wuhan.
The two colours represent places from China and rest of the world}
\label{fig3:casevsdist}
\end{figure}

We next study the spatial dependence of the infection spread.  The Haversine distances $d$ (in Km) between the  places 
of occurrence of the disease and  Wuhan are calculated and  
 the number of cases reported at a distance $d$ is  plotted in Fig. \ref{fig3:casevsdist}. 
For countries other than China, 
we take the distance from their capital cities to Wuhan.
The data of China alone can be   approximately fitted as   $R_{Ch}(d)  \propto d^{-2}$. 
However, the fit may not be appropriate when all the data are included as  the epicenters   might have shifted
elsewhere.  Indeed we find the data at larger $d$ to be more scattered. 
The correlation coefficient between $d$ and the affected number is also calculated; for China it is -0.267 while
for the rest of the world is is -0.197, agreeing with the above. 

\begin{figure}
\includegraphics[width = 7cm]{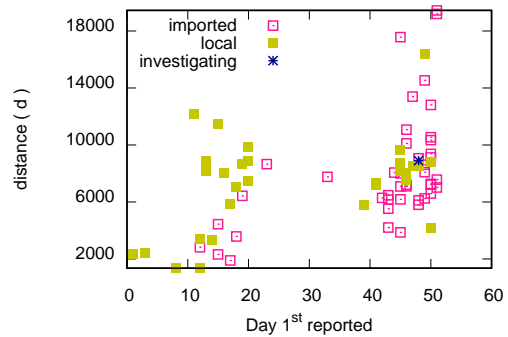}
\caption{The distance from Wuhan for the places of occurrence of the viral infection are plotted against the date of 
first occurrence ($t=0$  corresponds to January 13, 2020).}
\label{fig4:distvstime}
\end{figure}

For infection to other  countries,  there can be two modes of transmission: local or imported. 
We plot in Fig. \ref{fig4:distvstime} the distances from Wuhan against the 
date of the first reported cases in the countries in  the rest of the world (note that for this data, the 
origin of time is January 13, 2020 \cite{data}). 
We note that
there can be places at nearly the same distance which got the first infection on dates widely separated.  
However, in general, one can note  that the data points fall  around two clusters  quite well separated in time;
one dominated by infections due to  local transmission and the other consisting of  imported infections mainly, the latter occurring at later dates.   
%We also study the correlation of the first reported case date with the distance. 

%This system can
%describe various physical, biological or social problems e.g., pattern formation and opinion formation in
%society etc. 
%In the lattice representation of this model, 

To summarise, the space time dependence of the corona virus data show several interesting features as of now. 
The cumulative data for China show reasonably good agreement with the empirical form obtained in 
\cite{khaleq1}. Also, the numerical values of the timescales 
$T_{Ch}$ and $T_{ROW}$, 
associated with the  cumulative number of affected persons, 
are found to be quite close,   consistent with the fact that the data are for the same virus.  The epidemic is still in the rising stage outside China. The most intriguing result is the inverse square law dependence of the number of cases against distance from
the epicenter. This could have some connection with the gravity law in social dynamics.

\end{document}